\documentclass[twocolumn,superscriptaddress]{revtex4}

\usepackage{graphicx}

\begin{document}

\title{Impact of molecular structure on the lubricant squeeze-out
between curved surfaces with long range elasticity}

\author {U. Tartaglino}
\affiliation{IFF, FZ-J\"ulich, D-52425 J\"ulich, Germany}
\affiliation{International School for Advanced Studies (SISSA), and INFM
Democritos National Simulation Center, Via Beirut 2-4,  I-34014 Trieste,
Italy}

\author{I.M. Sivebaek}
\affiliation{IFF, FZ-J\"ulich, D-52425 J\"ulich, Germany}
\affiliation{Novo Nordisk A/S, Research and Development, DK-3400
Hiller{\o}d, Denmark}
\affiliation{MEK-Energy, Technical University
of Denmark, DK-2800 Lyngby, Denmark}

\author{B.N.J. Persson}
\affiliation{IFF, FZ-J\"ulich, D-52425 J\"ulich, Germany}

\author{E. Tosatti}
\affiliation{International School for Advanced Studies (SISSA), and INFM
Democritos National Simulation Center, Via Beirut 2-4,  I-34014 Trieste,
Italy}
\affiliation{The Abdus Salam International Centre for Theoretical Physics
(ICTP),  P.O.Box 586, I-34014 Trieste, Italy}

\date{\today}

\begin{abstract}

The properties of butane (C$_4$H${_{10}}$) lubricants confined
between two approaching solids are investigated by a model that
accounts for the curvature and elastic properties of the solid surfaces.
We consider the linear n-butane and the branched iso-butane.
For the linear molecule, well defined molecular layers develop in
the lubricant film when the width is of the order of a few atomic
diameters. The branched iso-butane forms more disordered structures
which permit it to stay liquid-like at smaller surface
separations.
During squeezing the solvation forces show oscillations
corresponding to the width of a molecule.
At low speeds ($<$ 0.1 m/s) the last layers of iso-butane
are squeezed out before those of n-butane.
Since the (interfacial) squeezing velocity in most practical
applications is very low when the lubricant layer has molecular thickness,
one expects n-butane to be a better boundary lubricant than
iso-butane.
N-butane possessing a slightly lower viscosity at high pressures,
our result refutes the view that squeeze out should be harder for
higher viscosities,
on the other hand our results are consistent with wear experiments in
which n-butane were shown to protect steel surfaces better than
iso-butane.

\end{abstract}

\maketitle

\section{\label{sec:intro}introduction}

Modern materials are subject to increasing loads and are used
under still more demanding conditions. This emphasises the need
for a better understanding of friction, lubrication and wear phenomena
\cite{P1,Israel}.

An example of the above is the fuel lubricated diesel engine
injection pump, capable of delivering in excess of 2000
bars today compared to 500 bars 15 years ago. This increase in
injection pressure has ensured higher engine efficiencies and lower
pollution levels.

Historically the diesel oil sulphur reduction both in the 1960s and
1990s combined with the high pressures have played a significant
role in injection pump durability. Sulphur, along with other
polarity inducing atoms, is part of polar species that ensure low
wear in boundary lubrication, a predominant regime in pumps
lubricated by diesel oil. Today there exist several accelerated laboratory
tests capable of predicting the lubricating abilities of a diesel
oil. If a fuel fails such a test, lubricity additives in small
proportions are added to ensure proper wear resistance.

\begin{figure}
\begin{center}
 \includegraphics[width=0.2\textwidth]{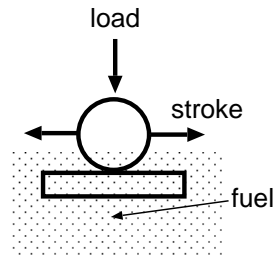} % fig_ballondisk.ps
\caption{\label{fig_ballondisk} The ball on disk configuration in a
HFRR wear test. }
\end{center}
\end{figure}

The currently used diesel oil wear test in Europe is the high
frequency reciprocating rig (HFRR) which is covered by several
standards \cite{hfrr_astm,hfrr_iso}. The principle in the HFRR is
the sliding of a fixed steel ball, loaded with 2N on a steel disk for 75
minutes. The motion is reciprocating with a stroke of 1 mm and a
frequency of 50 Hz. As the specimen contact is fuel lubricated the
resulting wear scar diameter on the ball expresses the lubricity of
the tested fuel. The configuration of the HFRR is shown in figure
\ref{fig_ballondisk}.

The HFRR ability of predicting the lifetime of injection pumps has
been questioned when low-viscosity fuels are tested. It appears that
a decrease in fuel viscosity requires an increase in lubricity
(smaller wear scar) to ensure full lifetime of the pumps. This
relation between viscosity and lubricity was confirmed as a new fuel
appeared in the 1990s: Dimethyl ether (DME). This fuel has a
viscosity 20 times lower than that of diesel oil and it was
established that an adequate lubricity level according to the HFRR
standards was not sufficient to protect the pump surfaces
\cite{mfprr2, vfvm, lub_exp_2003}.

\begin{figure}
 \includegraphics[width=0.475\textwidth]{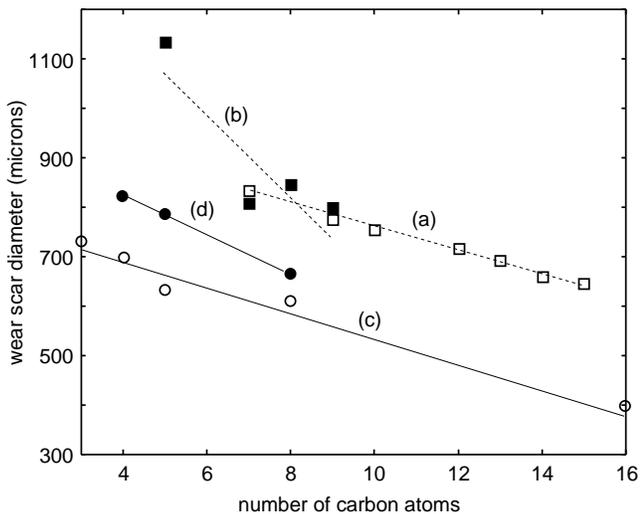} % fig_alkanes_2exp.eps
\caption{\label{fig_wsd_cn_2exp}The wear scar diameter for various
alkanes as a function of the molecular length. The squares are from
\protect\cite{weigaslub} and the circles from \protect\cite{sivsamper_2003}.
a), c) are for linear alkanes and b), d) for branched isomers.}
\end{figure}

A study \cite{sivsamper_2003} has argued that that fluid viscosity
is only a secondary property in the HFRR. Based on Molecular
Dynamics calculations it was discovered that the squeeze-out of
linear alkanes from surface contacts is primarily a function of the
length of the molecule. Experimental data shown in figure
\ref{fig_wsd_cn_2exp} demonstrates that branched alkanes perform
worse in the HFRR than their linear isomers although their
viscosities are almost the same \cite{lub_exp_2003}. This
observation confirms that the bulk viscosity may be a secondary
property in boundary lubrication.

The friction and wear properties in boundary lubrication of
lubricants with branched and linear molecules have been investigated
both experimentally \cite{hugo_sfa_n_i_octane, hugo_sfa_n_octane,
israel_branching_1989, Gee_cont_solid_1990} and using theoretical
simulation methods \cite{wang_md_i_n_octane_1994,
dijkstra_sim_n_i_alkanes_1997, tamura_md_thin_film_1999,
wang_md_branched_2002, cui_n_i_hexadecane_2001}.

Whether linear alkanes or their branched isomers are the best
lubricants according to the cited literature remains uncertain. A
number of studies indicate that n-alkanes are the best lubricants
\cite{wang_md_i_n_octane_1994,dijkstra_sim_n_i_alkanes_1997,cui_n_i_hexadecane_2001,wang_md_branched_2002}
whereas others claim that branched alkanes perform better
\cite{Gee_cont_solid_1990,hugo_sfa_n_i_octane,israel_branching_1989,tamura_md_thin_film_1999}.
One major weakness of the published studies is that the lubricant
squeezing is stopped at a separation of about eight \AA ngstrom
between the surfaces. At this point there are still a few monolayers
left protecting the surfaces, and significant wear originating from
cold welding should only appear when the last lubricant layer is
squeezed out \cite{Gee}.

In the present computer simulation study we investigate theoretically
the lubrication abilities of n-
and iso-butane extending the squeezing down to a surface separation
of zero \AA ngstrom so that the important expulsion of the last monolayer is
addressed.

\section{\label{sec:model}The model}
% The model: reference and short description. Maybe a picture.
% Features: elastic response for compression (Y) and shear deformation (G)
% Elastic energy important to get a reasonable description of the nucleation
% barriers.
% Curved surfaces; line contact.
% Picture with the sample
%
% Lubricant: United atom model, potential, potential for iso-butane
% Details: system size, number of atoms, lattice spacing (~ incommensurate).

\begin{figure}
 \includegraphics[width=0.475\textwidth]{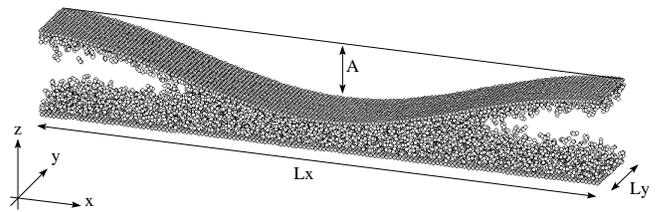} % s4n1.eps
 \caption{  \label{fig_a_sample}
 The simulated sample: two elastic walls with lubricant in between are
 pushed together. The sinusoidal profile of the upper surface generates
 a line contact in the middle, where lubricant forms a neck and eventually
 it is squeezed out by the load. Periodic boundary conditions are applied
 along $x$ and $y$.
 }
\end{figure}

\begin{figure}
 \includegraphics[width=0.475\textwidth]{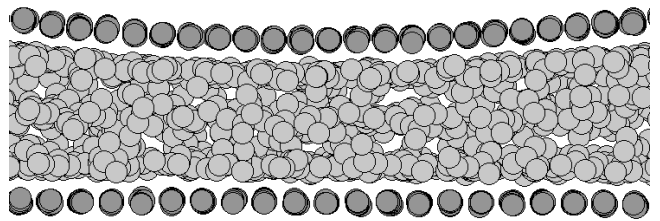} % s4i1_confined.eps
 \caption{ \label{fig_confined}
   The confined regime: the liquid lubricant get ordered into 2D
   layers when it is confined between atomically flat surfaces. The
   picture shows the side view of iso-butane in the contact region
   between the two walls. Light balls represent CH and CH$_3$ groups.
 }
\end{figure}

Figure \ref{fig_a_sample} shows the geometry for the simulation of the
squeeze-out process. Lubricant molecules are confined between two
elastic walls.  The lower wall, or {\em substrate}, is flat, while the
upper one, called {\em block}, has a sinusoidal profile, so that a
contact region appears in the centre of the sample. The liquid lubricant
adheres to the walls, forming a neck in the central region. When the
block is pushed towards the substrate, the lubricant has space to expand
laterally into the vapour. Periodic boundary conditions are applied
along the $x$ and $y$ directions. The periodically
repeated cell forms a rectangle $L_x\times L_y$,
with $L_x=416$~\AA\ and $L_y=52$~\AA. The contact region is
not circular, but forms an infinite long strip parallel to the $y$
axis.

The lubricant molecules are described through the Optimized Potential
for Liquid Simulation (OPLS) \cite{jorgensen1984x1,dysthe2000x1}; this
potential is known to provide density and viscosity of hydrocarbons
close to the experimental one.

Each butane molecule comprises four units (particles), each particle
corresponding to one chemical group CH$_3$, CH$_2$ or CH ({\em United
Atom} model). The interaction between particles of different molecules
is described by Lennard-Jones potentials.  The intramolecular interactions
include two body forces that keep the bond length C--C close to 1.53
\AA ngstrom, three body forces imposing a preferred angle of 115 degrees
between the carbon atoms, and four body forces favouring a well defined
torsion of the molecules.  The four body forces apply to the sequence of
carbon atoms C--C--C--C, thus they are present only for n-butane.

For the iso-butane molecules we introduced also an anti umbrella
inversion potential, analogously to what already done by Mondello and
Grest \cite{mondello_grest} for branched hydrocarbons; the only
difference here being that in the iso-butane molecule it is not possible
to distinguish which carbon atom constitutes the lateral branch and
which constitute the backbone of the molecule. Therefore we symmetrized
the anti inversion potential.

The interaction between the beads of the lubricant molecule and the
walls' atoms is also given by Lennard-Jones forces, whose energy parameter
$\epsilon_0=18.6$ meV, as in ref.~\cite{xia1992x1}, yielding an adsorption
energy of about 0.3 eV per molecule, comparable to that of butane on
Au(111) \cite{wetterer}.
% Some rough estimate from my simulations:
% Adsorption of single molecules
%   E_adsorption = 0.274 eV/molecule (n-butane)
%                = 0.271 eV/molecule (iso-butane)
% Adsorption of 2-layer-thick film
%   E_adsorption = 0.333 eV/molecule (n-butane)
%                = 0.323 eV/molecule (iso-butane)

The elastic energy due to the walls' deformation is also involved in
the nucleation process that triggers the transition from $n$ to $n-1$
monolayers of lubricant in the confined regime \cite{pertos1994}.
In principle we can achieve a realistic description of the walls'
elastic response by simulating thick walls comprising many layers.
Since we necessitate proper elastic behaviour up to wavelengths
comparable with the size of the contact area the walls' thickness must
not be smaller that the lateral size of the box, $L_y=52$~\AA.
This indeed would imply too many atoms, slowing down the calculations.
Instead we adopted the model described in ref.~\cite{perbal_2000}: for
each wall only the outermost layer of atoms is considered; these atoms
are connected to a rigid surface with springs that take into account
both the compressibility and the shear rigidity of the wall. Similar
springs connect together the square grid of atoms of the wall.
The advantage of this approach is double: with a relatively small number
of atoms it is possible to describe the long range elasticity of the
walls, and to impose a curved profile simply by using a curved rigid
surface.

The rigid surface connected to the substrate's atoms is flat and its
position is fixed. The upper rigid surface has the profile
\begin{equation}
  \label{z_upper}
  z(x,y) = z_0 + \frac{A}{2} \left( 1 - \cos \left( \frac{2 \pi x}{L_x}
  \right) \right) \, ,
\end{equation}
where the amplitude of corrugation $A$ is 20, 40, or 60 \AA. The case
$A=20$ \AA\ guarantees a larger contact region, but it does not allow
enough empty space (or rather vapour space region)
for the squeeze-out of the last two layers; it was used
only for comparison with the other cases, to ensure that the size of the
contact region is large enough for the nucleation of the squeeze-out
process.

The springs connecting the walls' atoms to the rigid surfaces simulate
the elastic response of a gold film of thickness 50 \AA ngstrom,
comparable with the lateral size $L_y$ of the sample.

The substrate consists of $144 \times 18$ atoms in a square lattice with
lattice spacing 2.889 \AA ngstrom; the block layer is made of $160\times
20$ atoms, with lattice spacing 2.6 \AA ngstrom. The different lattice
spacing is to reduce the commensurability between the walls. Both
substrate's and block's atoms have the same mass of gold: 197 a.m.u.
In most of the simulations we employed 2800 lubricant molecules, which
deposit on the two walls forming about two monolayers of adsorbate
on each surface.

\section{\label{sec:results}Simulation results}
% Squeeze of the thick layers (4->3->2->1) show that iso-butane is removed
% more easily.
% Picture with pressure vs time (S4, T=300 K, h=40 Ang, v=1 m/s)
% The last layer of iso-butane is harder to remove. Results for the last
% layer (S5, T=300 K, h=60 Ang, v=1 m/s).
% Picture with the snapshots versus time of the lubricant in the contact
% region
% Thermodynamic of kinetic effect? Even at higher temperatures
% the result is the same (S9, T=330 K) (S8, T=350 K).
% Picture with pressure vs time (S8, T=350) showing the squeeze-out up to
% the latest monolayer. The plateau for iso-butane.
% Squeezing slowly enough reveals that the iso-butane lubricity was a
% kinetic phenomenon. Even at v=0.1 iso-butane still appears more resistent,
% with v=0.03 it has enough time to overcome the nucleation barriers
% before n-butane.
% Picture: pressure versus time (S10, T=300 K, v=0.03; or S13 or S14).
% We observed that lateral sliding tend to reduce the kinetic factor,
% but we do not have enough statitistics for a strong claim.

\begin{figure}
 \includegraphics[width=0.475\textwidth]{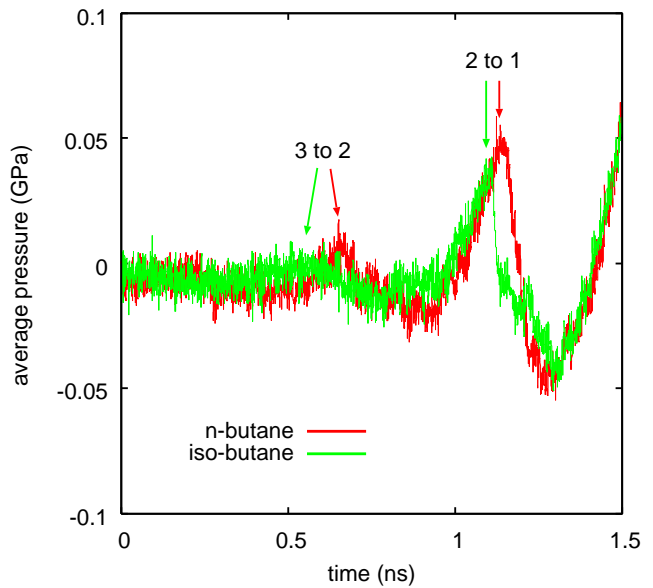} % s4_pressure.eps
 \caption{  \label{fig_pressure_vs_time_300K}
 Squeeze-out of butane at $T=300$ K\@. The pressure drops when the
 number of layers switches from 3 to 2 and from 2 to 1.
 Squeezing speed 1 m/s.
 }
% T=300 K; Vz=-1 m/s; A=40 Ang; 3->2->1, iso-butane first; no more vacuum.
% The high pressure region is 15% of the whole area.
\end{figure}

\begin{figure}
 \includegraphics[width=0.475\textwidth]{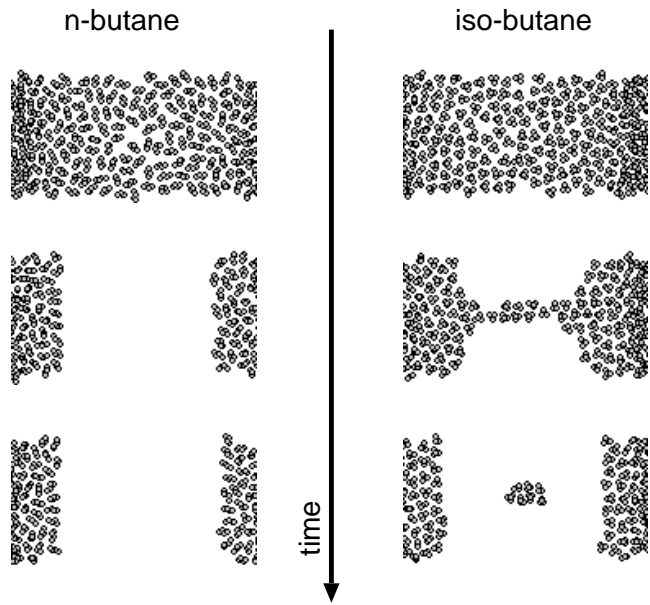} % s5_lastlayer.eps
 \caption{  \label{fig_lastlayer_300K}
 Top view of the lubricant molecules in the central part of
 the contact area immediately
 before the expulsion of the last monolayer (top), after 0.5 nanoseconds
 (middle), and after 1 nanosecond. $T=300$ K, squeezing speed 1 m/s.
 (Block and substrate are parallel to the plane of the figure)}
% T=300 K; Vz=-1 m/s; A=60 Ang; 1->0 iso-butane first.
% Snapshots of the contact region 0.5 ns before, during,
% and 0.5 ns after the squeeze-out.
\end{figure}

% Pushing down at constant speed, with identical conditions for iso- and
% n-butane
Two simulations are prepared in parallel for iso- and n-butane with
identical conditions. Initially the whole system is thermalized and
the lubricant adheres to the two walls. Then the rigid upper surface
is moved down towards the substrate with constant speed, i.e., $z_0$
in equation \ref{z_upper} is constrained to decrease linearly with
time. As the lubricant layers on the two walls
get in contact, a neck is formed in the contact area. The lubricant
clearly shows layering both for the n-butane and for the more
disordered iso-butane (Fig.~\ref{fig_confined}). The change in the
number of monolayers takes place abruptly, with a sudden decrease of
pressure due to the relaxation of the elastic energy stored in the
walls.

% iso-butane is squeezed out more easily, except when the last layer
% goes away.
Fig.~\ref{fig_pressure_vs_time_300K} shows the behavior of pressure versus time
while squeezing at speed 1 m/s and at temperature 300 Kelvin. The
$y$-axis contains a spatial average of the pressure, that is the
vertical force divided by the area $L_x L_y$ of the simulated cell.
The contact region is about 15\% of the whole cell size,  and the real
pressure in the center of the contact area is much larger than the average
pressure.
Negative pressures observed in some time intervals
are due to the attractive capillarity forces of the neck of lubricant.
The plotted curves show that iso-butane is squeezed out earlier and at a
lower pressure. This is the typical behaviour we observed when there is
more than one monolayer. Conversely the removal of the very last
monolayer appears to be easier for n-butane. As can be observed in
Fig.~\ref{fig_lastlayer_300K}, the ejection of the iso-butane not only
happens later, but it is slower and a small group of 7 molecules remains
trapped at the end. This suggests that the simulated squeeze-out of the last layer
is strongly affected by dynamical effects, such as the drag friction when
the lubricant slides on the solid walls. Thus, even though the nucleation barrier of
the hole may be smaller for iso-butane, the simulated system might not have enough time
to exit its metastable state before the pressure is further increased
by the vertical (squeezing) motion of the block.

\begin{figure}
 \includegraphics[width=0.475\textwidth]{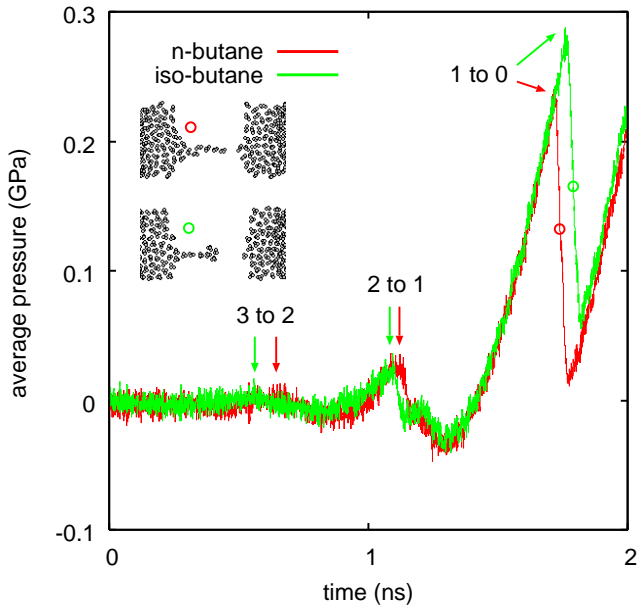} % s8_pressure.eps
 \caption{  \label{fig_pressure_vs_time_350K}
 Pressure versus time during the complete squeeze-out of three
 monolayers at $T=350$ K, squeezing speed 1 m/s. The pictures show the
 lubricant while the last monolayer is being removed.
 Corrugation of the upper profile: $A=60$ \AA; contact area: 10\%
 of the cell's area.
% T=350 K; Vz=-1 m/s; A=60 Ang; 3->2->1 iso-butane first; 1->0 n-butane 1st
% The simulations S5 (T=300 K) and S9 (T=330 K) show the same: 1-0 n-b 1st.
% The high pressure region is 10% of the whole area.
 }
\end{figure}

\begin{figure}
 \includegraphics[width=0.475\textwidth]{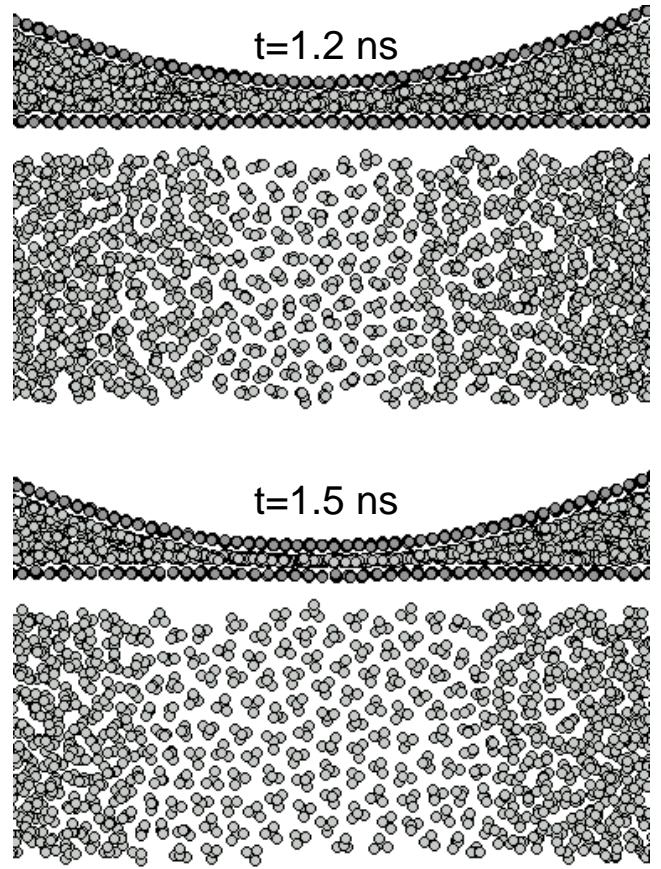} % plateau.eps
 \caption{  \label{fig_plateau}
 Side and top views of the iso-butane monolayer at different time (and
 pressure). Images refer to the simulation of Figure
 \protect\ref{fig_pressure_vs_time_350K}.
 }
\end{figure}

% At higher temperature it is the same
The behaviour at higher temperatures is qualitatively the same. At
$T=350$ Kelvin the switch from 3 to 2 monolayers and from 2 to 1
monolayer happens first for iso-butane, but the very last layer of
iso-butane stays longer between the walls, and 5 iso-butane molecules
remain trapped there (see Fig.~\ref{fig_pressure_vs_time_350K}).
% The real pressure in the contact
% region is much larger than the average value shown along the
% $y$-axis (the contact region is one tenth of the whole size, and the
% corrugation of the upper profile is $A=60$ \AA\ to provide enough space
% for the complete squeeze-out).
Another independent simulation at $T=330$ K (not shown) confirms this
trend.

% The plateau: two stages transition from 2 to 1 monolayer of
% iso-butane.
The transition from 2 to 1 iso-butane monolayer reveals another
interesting feature: the pressure drops down in two stages, as is
shown in the plot of figure \ref{fig_pressure_vs_time_350K} for time
close to 1.2 ns. What happens is clearly illustrated in the snapshots
of figure \ref{fig_plateau}: initially there is the squeeze-out of one
of the two layers. The remaining monolayer does not have a preferential
orientation of the molecules parallel to the walls, the plateau in the
pressure versus time graph at $t\approx 1.2$ ns is due to this {\it thick}
monolayer. Finally the molecules change their orientation
forming a {\it thinner} monolayer, yielding a smaller second pressure drop.
This kind of two-stage squeezing of the second layer has
been observed in other simulations too, but only for iso-butane. It is
in fact a steric effect due to the shape of the molecule.
Similar pressure-induced {\it phase transitions} have been observed in
other computer simulations of squeeze-out.
Thus, in Ref.~\cite{squeeze_out_review}
it was shown that Xe atoms between two solid surfaces
exhibited triangular bilayers at low squeezing pressure,
which abruptly transformed to
fcc(100) layers parallel to the solid surfaces when the pressure increased.
This transition allowed the surfaces to move closer to each
other which released elastic energy.

\begin{figure}
 \includegraphics[width=0.475\textwidth]{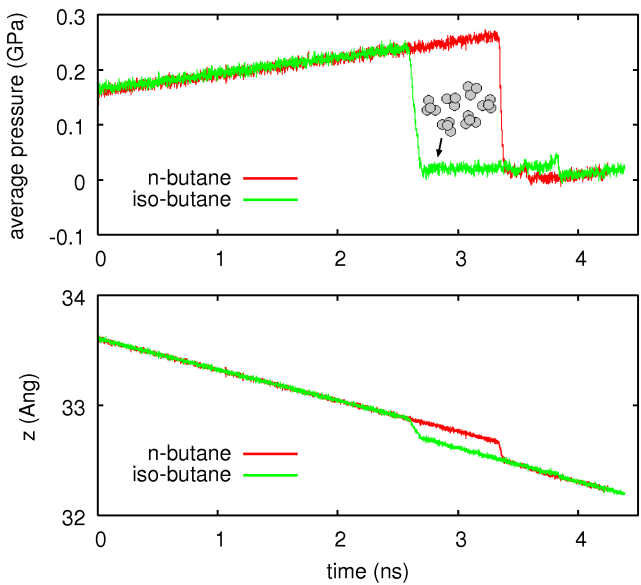} % s10_lowspeed.eps
 \caption{  \label{fig_pressure_vs_time_slow}
 Last monolayer squeeze-out at low speed. $T=300$ K, squeezing speed
 0.03 m/s. Iso-butane is removed earlier at a lower pressure, but the
 cluster of six molecules shown in the picture remains trapped up
 to $t=3.6$ ns.
 Below: the average position of the upper wall's atoms versus time.
 }
% T=300 K; Vz=-0.03 m/s; A=60 Ang; 1->0 n-butane 1st
% But at Vz=-0.1 m/s (S6) or at higher temperatures (T=350 K, S14) iso-b 1st.
% At T=330 K (S13) same time.
% The high pressure region is 10% of the whole area.
\end{figure}

\begin{figure}
 \includegraphics[width=0.475\textwidth]{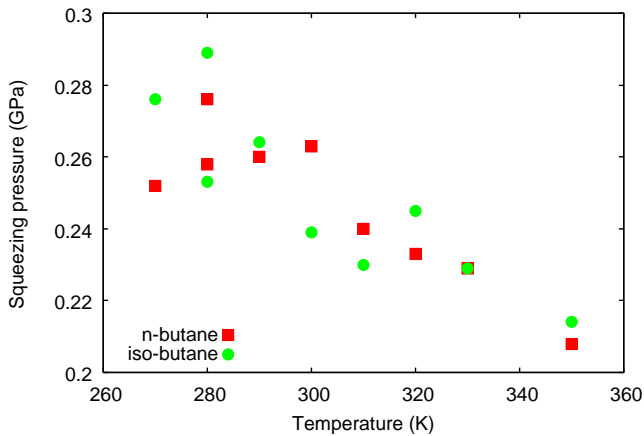} % squeeze_vs_T.eps
  \caption{  \label{fig_squeeze_vs_T}
   Pressure required to squeeze-out the last lubricant layer
   while pushing at speed 0.03 m/s. Data refers to the spatial
   average over the whole simulation cell. The pressure in the centre
   of the contact area is much larger. Each point is the result of a single
   independent simulation, requiring about 2 mounths of CPU-time on a
   standard PC.
  }
\end{figure}

% Pushing slowly
The simulations at lower squeezing speed give clear evidence that the
the relatively higher lubricity of the last layer of iso-butane is a
kinetic effect. Thus,  at $T=300$ K and for the compression
speed 1 m/s (Fig.~\ref{fig_lastlayer_300K})
n-butane is removed first, while when the speed is 0.1 m/s the removal of the two
lubricants happens almost at the same time (not shown). Finally when the
squeezing is carried at the speed 0.03 m/s iso-butane is removed first
and at a lower pressure (Fig.~\ref{fig_pressure_vs_time_slow}). In this
case there is still some small island of molecules trapped, shown in the
picture, confirming once more that the lateral sliding of iso-butane
is slower.

% Squeezing pressure versus speed of pushing
A further evidence that the removal of the last monolayer is influenced by
kinetic effects comes from the squeezing pressure versus pushing speed
$v_z$: keeping all the other conditions identical, we observed that the
pressure at which the last layer is removed does decrease when the block
is pushed down slowly. For example, for iso-butane at $T=300$ K we
obtained the following pressures: 0.340 GPa when $v_z=1$ m/s, 0.257 GPa
when $v_z=0.1$ m/s, and 0.239 GPa when $v_z=0.03$ m/s (as usual, these
are spatial averages, the true pressures in the contact area being much
higher).  Similarly, the corresponding simulations for n-butane
gave 0.321, 0.252, and 0.263 GPa respectively. The lubricant is able to
nucleate the hole for the squeeze-out at the lowest of these pressures,
but at high pushing speed there is not enough time to exploit this
possibility before the pressure is further increased.
% Squeezing pressure for the last monolayer versus speed (T=300K):
% Speed v_z (m/s)   n-butane (GPa)   iso-butane (GPa)
%  0.03              0.263            0.239
%  0.10              0.252            0.257
%  1.00              0.321            0.340

% The trend with temperature
It is interesting to observe what happens when the temperature is
changed, particularly considering that in reality the microcontacts
where the squeeze-out happens are likely to be much warmer than the
environment, at least during sliding.
The temperature is likely to speed up the mobility of the
molecules, reducing the influence of the drag force of the walls.
Moreover the larger thermal fluctuations should strongly favour the
nucleation of the critical hole in the lubricant layer. Both effects
tend to reduce the pressure needed to squeeze the lubricant. This is
indeed the trend which we found and that is observed in
Fig.~\ref{fig_squeeze_vs_T}.
On the other hand, any difference $\Delta G$ between n-butane and
iso-butane in the free energy barriers for nucleation of squeeze-out
should become less important as the temperature increases, as the
ratio between the nucleation probabilities is mainly affected by the Boltzmann factor
$\exp(-\beta \Delta G)$, which goes to 1 when $T\to\infty$.

Unfortunately the low temperature conditions are hard to analyse through
simulations: the randomness of the single squeeze-out event is enhanced
by the strong decrease of the Boltzmann factor. Actually the transition
time for a nucleating process has an exponential distribution,
where the standard deviation is equal to the average value.
Thereafter the data of a single simulations become no more reliable
and only an average over many independent simulations ---not easily
affordable due to computer time limitations--- would show the real trend.
We do not have enough statistics in Fig.~\ref{fig_squeeze_vs_T}
for a reliable comparison of the behaviours of n-butane and iso-butane
below room temperature.

% Lateral sliding
Finally we consider the effect of lateral sliding of the walls with
respect to each other. We ran some simulations (not shown) at
different temperatures with squeezing speed and lateral speed both
equal to 1 m/s, and we observed that the removal of the last
monolayer happens almost simultaneously at the same pressure. The
relative motion of the walls tends to take away the effect of the
drag between walls and lubricant, which is indeed responsible of the
higher lubricity of iso-butane when squeezing fast. This is an
important remark, since in many practical situations there is
lateral sliding.

% Implications for experimental work added by IMS 26 October 2005.

The significance of the temperature for the squeeze-out of the two
different butanes is of outermost importance for future wear
experiments. The results in figure \ref{fig_wsd_cn_2exp} are all
obtained at 298K. At this temperature and at the given sliding
conditions it seems that iso-butane is more easily squeezed out from
the ball-disk contact than n-butane as the resulting wear is
slightly lower for the latter one. In real injection pumps the
temperature could be very low (at start up) and very high (running
conditions). The present simulation results imply that wear
protecting properties of different molecules in the fuel may not
react the same way to changes in temperature and sliding conditions.
This means that a wear test may not reflect reality unless test
conditions such as temperature and sliding velocity are varied. This
new aspect of wear experiments should be looked into in the near
future.

\section{\label{sec:conc}summary and conclusions}

We have studied the properties of butane (C$_4$H${_{10}}$) lubricants confined
between two approaching solids using a model that
accounts for the curvature and elastic properties of the solid surfaces.
We considered both linear n-butane and the branched iso-butane. In the
case of the linear molecule well defined molecular layers develop in
the lubricant film when the width is of the order of a few atomic
diameters. The branched iso-butane forms more disordered structures
which permit it to stay liquid-like at smaller surface
separations.

During squeezing the solvation forces show oscillations
corresponding to the width of a molecule.
At low speeds ($<$ 0.1 m/s) the last layers of iso-butane
are squeezed out before those of n-butane.
Since the (interfacial) squeezing velocity in most practical
applications is very low when the lubricant layer has molecular thickness,
one expect n-butane to be a better boundary lubricant than
iso-butane.
This is consistent with wear experiments in
which n-butane was shown to protect steel surfaces better than
iso-butane.
On the other hand, n-butane possessing lower viscosity at high
pressures, our result refutes the view that squeeze out should be
harder for higher viscosities.
At high squeezing velocity, the squeeze-out of the last monolayer of
iso-butane occurs at higher pressures than for n-butane. We
interpret this as a kinetic effect resulting from the lateral
corrugation barrier experienced by the molecules. As the temperature
increases, the squeeze-out occurs at lower applied pressure for both
n-butane and iso-butane. This is the expected result based on the
nucleation theory of
squeeze-out\cite{pertos1994,squeeze_out_review}.

\begin{acknowledgments}

U.T. and I.M.S. acknowledge support from IFF, FZ-J\"ulich,
hospitality and help of the staff during their research visits.
I.M.S. also acknowledges financial support from the European project
AFFORHD. In Trieste, this work was also sponsored by
MIUR COFIN No.\ 2003028141-007, MIUR COFIN No.\ 2004028238-002,
MIUR FIRB RBAU017S8 R004, and MIUR FIRB RBAU01LX5H.

\end{acknowledgments}

\end{document}